\documentclass[prb,superscriptaddress,preprint,endfloats,showpacs]{revtex4}
\usepackage{amsmath, amsthm, amssymb}
\usepackage{graphicx}
\usepackage{dcolumn} 
\usepackage{bm} 
\makeatletter 
\gdef\@ptsize{2}
\let\@currsize\normalsize 
\makeatother 
\usepackage{setspace} 

\newcommand {\BIO}{Ba$_{2}$IrO$_{4}$}
\newcommand {\SIO}{Sr$_{2}$IrO$_{4}$}
\newcommand {\PSO}{PrScO$_{3}$}
\newcommand {\LTO}{LaTiO$_{3}$}
\newcommand {\zerozero}{$(0,0)$}
\newcommand {\pizero}{$(\pi,0)$}
\newcommand {\pipi}{$(\pi,\pi)$}
\newcommand {\halfhalf}{$(\pi /2,\pi /2)$}
\newcommand {\EF}{$E_{\mathrm{F}}$}
\newcommand {\ttwog}{$t_{2\mathrm{g}}$}
\newcommand {\Jhalf}{$J_{\mathrm{eff}}=1/2$}
\newcommand {\Jeff}{$J_{\mathrm{eff}}$}

\newcommand {\Jthreehalf}{$J_{\mathrm{eff}}=3/2$}
\newcommand {\TC}{$T_{\mathrm{c}}$}
\newcommand {\TN}{$T_{\mathrm{N}}$}
\usepackage[usenames,dvipsnames]{color}

\begin{document}
\title{Correlated vs. conventional insulating behavior in the {\Jhalf} vs. $3/2$ bands in the layered iridate {\BIO}}  

\author{M. Uchida}
\altaffiliation[Present address: ]{Department of Applied Physics and Quantum-Phase Electronics Center (QPEC), University of Tokyo, Tokyo 113-8656, Japan}
\affiliation{Laboratory of Atomic and Solid State Physics, Department of Physics, Cornell University, Ithaca, New York 14853, USA}
\author{Y.F. Nie}
\affiliation{Laboratory of Atomic and Solid State Physics, Department of Physics, Cornell University, Ithaca, New York 14853, USA}
\affiliation{Department of Materials Science and Engineering, Cornell University, Ithaca, New York 14853, USA}
\author{P.D.C. King}
\affiliation{Laboratory of Atomic and Solid State Physics, Department of Physics, Cornell University, Ithaca, New York 14853, USA}
\affiliation{Kavli Institute at Cornell for Nanoscale Science, Ithaca, New York 14853, USA}
\author{C.H. Kim}
\affiliation{Department of Applied Physics, Cornell University, Ithaca, New York 14853, USA}
\author{C.J. Fennie}
\affiliation{Department of Applied Physics, Cornell University, Ithaca, New York 14853, USA}
\author{D.G. Schlom}
\affiliation{Department of Materials Science and Engineering, Cornell University, Ithaca, New York 14853, USA}
\affiliation{Kavli Institute at Cornell for Nanoscale Science, Ithaca, New York 14853, USA}
\author{K.M. Shen}
\email[Author to whom correspondence should be addressed: ]{kmshen@cornell.edu}
\affiliation{Laboratory of Atomic and Solid State Physics, Department of Physics, Cornell University, Ithaca, New York 14853, USA}
\affiliation{Kavli Institute at Cornell for Nanoscale Science, Ithaca, New York 14853, USA}

\date{\today}


\begin{abstract}

We employ molecular beam epitaxy to stabilize {\BIO} thin films and utilize {\textit{in situ}} angle-resolved photoemission spectroscopy to investigate the evolution of its electronic structure through the N{\' e}el temperature $T_{\mathrm{N}}$. Our measurements indicate that dispersions of the relativistic {\Jhalf} and 3/2 bands exhibit an unusual dichotomy in their behavior through the N{\' e}el transition. Although the charge gap survives into the paramagnetic state, only the {\Jhalf} state exhibits a strong temperature dependence and its gap softens with increasing temperature approaching $T_{\mathrm{N}}$, while the nearly fully occupied {\Jthreehalf} state which remains nearby in energy exhibits negligible changes with temperature.

\end{abstract}
\maketitle

\section{Introduction}

$5d$ transition metal oxides have recently attracted great interest due to the interplay between spin-orbit and Coulomb interactions which can give rise to novel many-body quantum states, including the theoretically proposed Weyl semimetal,\cite{SOIweyl1, SOIweyl2} topological Mott insulator,\cite{SOItopo1, SOItopo2} or high-temperature superconductivity upon carrier doping.\cite{iridateSC, iridateSC2TB2} These studies have been largely motivated by the experiments in a model compound {\SIO}, which suggest that the low-energy electronic states can be suitably represented by their effective total angular momentum {$J_{\mathrm{eff}}$}, and argue that the $\mathrm{Ir}^{4+}$ ($5d^{5}$) {\ttwog} orbitals are split into a fully filled {\Jthreehalf} manifold and a singly occupied {\Jhalf} band.\cite{iridateSOI1, iridateSOI2} It has been proposed that the modest Coulomb interactions of the $5d$ electrons can then result in a further splitting of the half-filled {\Jhalf} band into a fully occupied valence band and an unoccupied conduction band.

At present, there remains considerable debate about whether iridates are better described as Mott insulators,\cite{SIOopt1} in analogy to the conventional $3d$ transition metal oxides, or as Slater insulators,\cite{slater1, slater2, slaterSTM} where the insulating behavior is tied directly to long-range antiferromagnetic order. For example, recent scanning tunnel spectroscopy on {\SIO} has reported an onset of the temperature dependent spectra below the antiferromagnetic ordering temperature $T_{\mathrm{N}}$, consistent with dynamical mean field theory results,\cite{slaterSTM} although the in-gap spectral weight is still strongly suppressed even above $T_{\mathrm{N}}$, as also seen in other spectroscopy experiments.\cite{SIOopt1, slater1} In order to definitively address these issues, it has become critical to systematically examine other related $5d$ transition metal oxides, such as {\BIO}. This also opens new avenues to realize proposed exotic phases such as superconductivity which may exist at the intersection between strong spin-orbit interactions and electron correlations. 

\begin{figure}
\begin{center}
\includegraphics*[width=13cm]{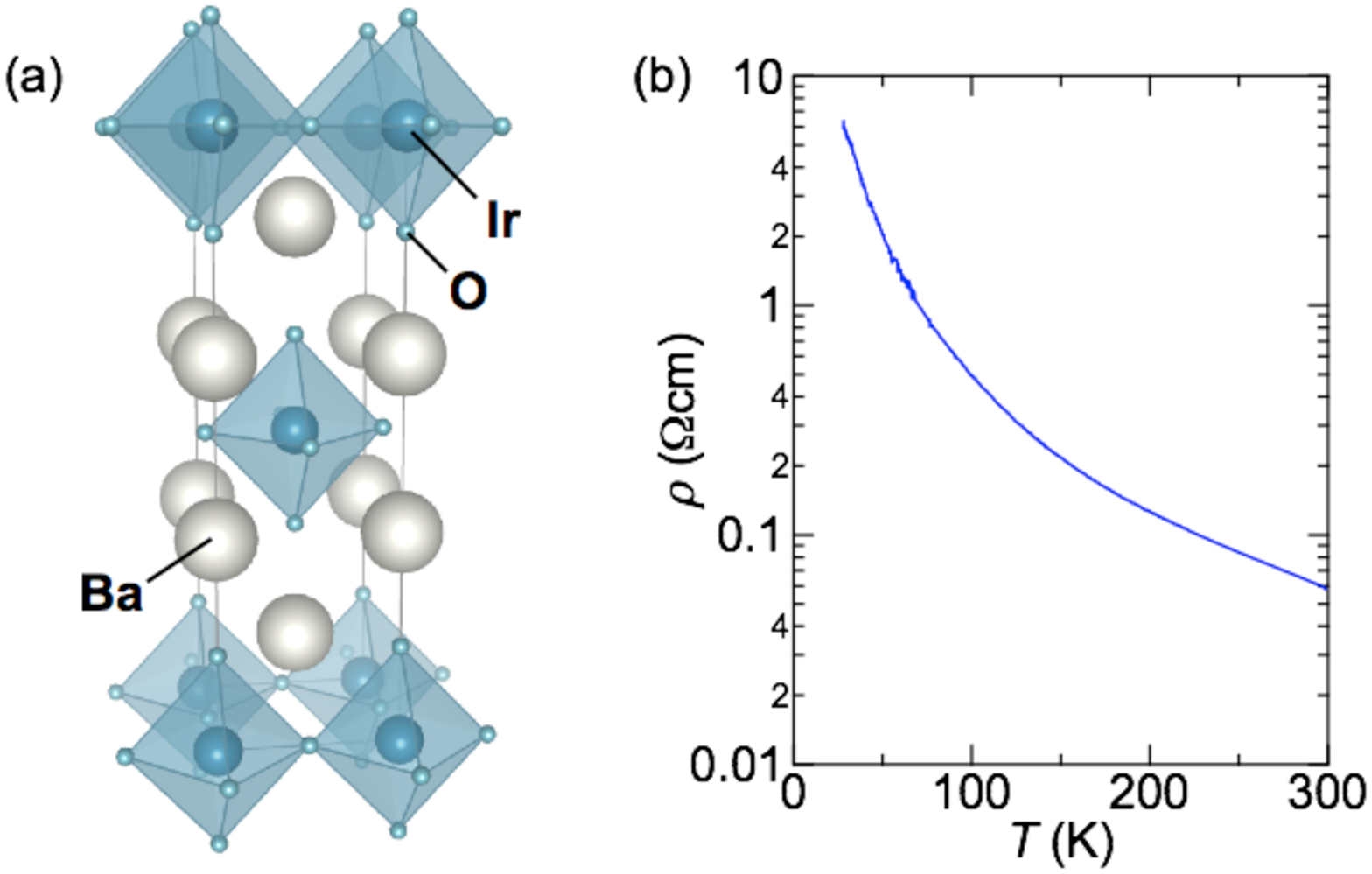}
\caption{(Color online) (a) Structure view of {\BIO}. (b) Temperature dependence of the longitudinal in-plane resistivity of a 16 nm {\BIO} film grown on {\PSO}.
}
\label{fig1}
\end{center}
\end{figure}

{\BIO} is an ideal candidate material for addressing many of the outstanding issues in the field of the iridates. {\BIO} possesses a less distorted, simple quasi-two-dimensional crystal structure (\textit{I}4/\textit{mmm} in bulk), as shown in Fig. 1(a), with Ba replacing Sr,\cite{BIO1} while it shows the same insulating behavior accompanied with the basal plane antiferromagnetic order below the N{\' e}el temperature $T_{\mathrm{N}}=240$ K.\cite{BIO3RXS} The undistorted crystal structure without the in-plane IrO$_6$ octahedral rotation is expected to result in simplified band dispersions without folded features, enabling us to follow the detailed temperature evolution of the relativistic {\Jhalf} and 3/2 bands and to determine their roles in the formation of the insulating phase. In addition, {\BIO} has a number of distinct advantages over {\SIO}, including the capacity for being metallized either through carrier doping (K or La substitution for Ba) \cite{BIOdoping} or through the application of hydrostatic pressure,\cite{BIO2} making it an excellent platform for searching for the possibility of exotic, iridate-based superconductivity. Unfortunately, layered {\BIO} is metastable in bulk and can only be formed under high pressure, making the synthesis of large bulk single crystals difficult. Here, we apply oxide molecular beam epitaxy (MBE) to stabilize thin films of the desired structure of {\BIO}, and utilize {\textit{in situ}} angle-resolved photoemission spectroscopy (ARPES) to investigate the parent insulating ground state of {\BIO} and how it evolves through $T_{\mathrm{N}}$. Longitudinal in-plane resistivity in Fig. 1(b) shows clear insulating characteristics of the {\BIO} film. We compare our ARPES measurements with calculations based on density functional theory incorporating electron-electron and spin-orbit interactions. While both the relativistic {\Jhalf} and 3/2 states are nearly degenerate near {\EF}, we reveal a surprising contrast in their temperature dependence. Although the gap survives well into the paramagnetic state, only the {\Jhalf} band exhibits a substantial broadening and `softening' of the gap with increasing temperature approaching $T_{\mathrm{N}}$ reminiscent of some of the parent cuprate superconductors,\cite{cuprateFC} suggesting the importance of long-range antiferromagnetic order. 

\section{Experimental and calculational procedures}

Thin films of (001) {\BIO} of thickness $\sim$15 nm were deposited epitaxially on (001)$_{p}$ (where the subscript $p$ denotes pseudocubic indices) {\PSO} single crystal substrates using a Veeco GEN10 oxide MBE system. Absorption-controlled deposition was performed in distilled 100\% $\mathrm{O}_{3}$ at a background pressure of $1\times10^{-6}$ Torr. Under optimized adsorption controlled conditions, Ba and Ir were supplied continuously with a flux of $6\times10^{12}$ and $7\times10^{12}$ $\mathrm{atoms}/\mathrm{cm}^2 \mathrm{s}$ from an effusion cell and an electron beam evaporator, respectively. The substrate was heated at 800 $^{\circ}$C as measured by a pyrometer. Under these conditions the extra Ir atoms form volatile $\mathrm{IrO}_{x}$ \cite{IrOx} and then evaporate from the surface, leaving stoichiometric {\BIO} films.

During growth films were monitored with reflection high-energy electron diffraction (RHEED) while rotating the substrate. After growth, samples were immediately ($<$ 300 seconds) transferred through ultrahigh vacuum to a high-resolution ARPES system. Measurements were performed using a VUV5000 helium plasma discharge lamp and a VG Scienta R4000 hemispherical electron spectrometer, with an excitation energy of 21.2 eV (He I$\alpha$) and an instrumental energy resolution of 20 meV. The sample was measured at temperatures between 100 and 300 K while maintaining a base pressure typically better than $8 \times10^{-11}$ Torr. Following ARPES measurements, films were characterized {\textit{in situ}} by low-energy electron diffraction (LEED). The phase purity and crystallinity of {\BIO} films were characterized also using {\textit{ex situ}} four-circle x-ray diffraction (XRD) with Cu $K\alpha$ radiation. 

The {\textit{ab-initio}} density functional theory (DFT) calculations were performed using Wien2k code \cite{wien2k} including spin-orbit coupling and an on-site Coulomb repulsion, with the local density approximation and the Perdew-Burke-Ernzerhof exchange-correlation functional.\cite{PBE} A tetragonal structure was assumed, with the slightly strained lattice ($a=4.021$ {\AA} and $c=13.34$ {\AA}) on the {\PSO} substrate, and no peaks corresponding to octahedral rotation were observed from bulk diffraction measurements.

\section{Film characterization}

\begin{figure}
\begin{center}
\includegraphics*[width=15cm]{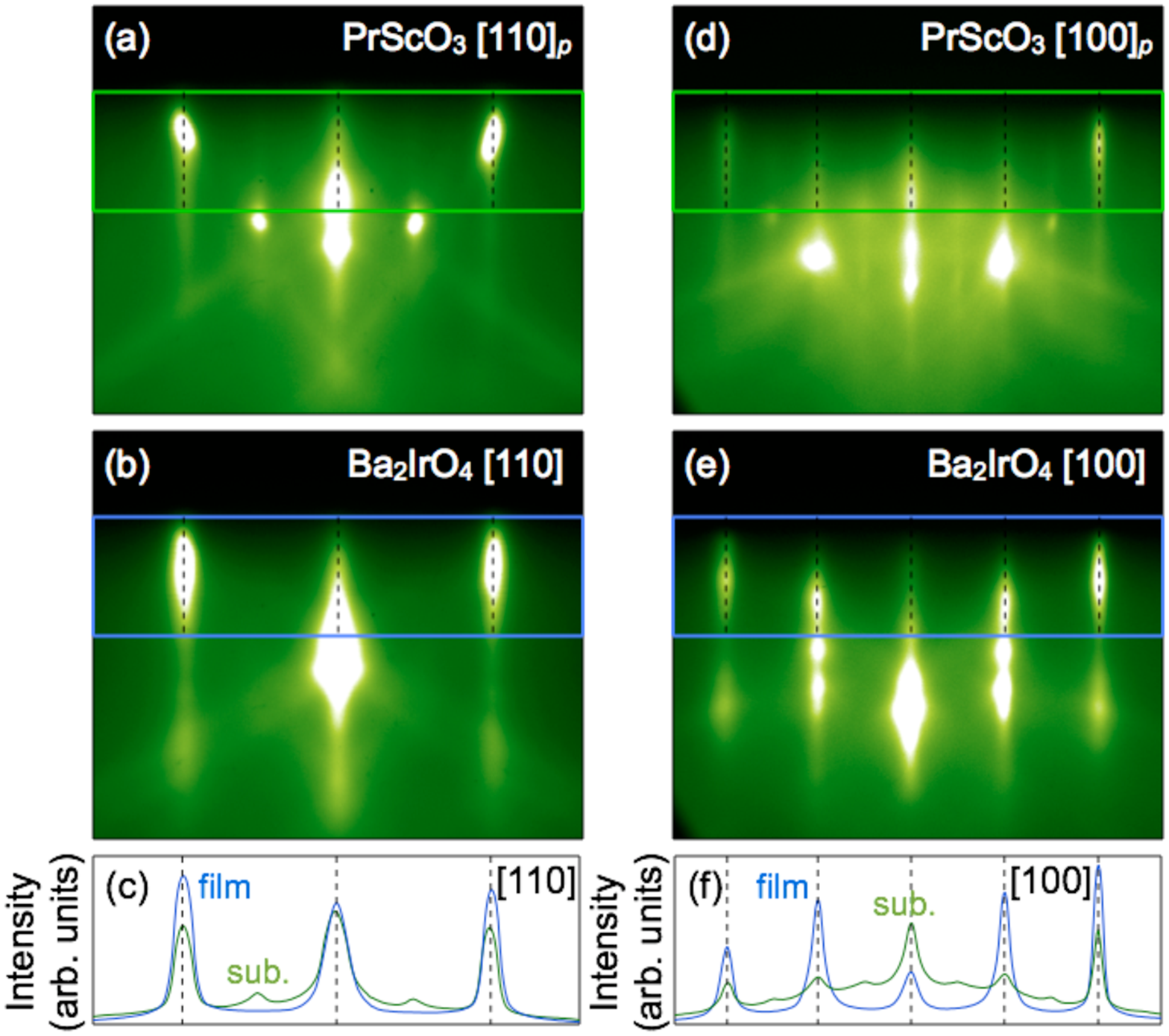}
\caption{(Color online) RHEED images of (a) the bare {\PSO} substrate and (b) after the growth of a 16 nm thick {\BIO} film, taken along the [110]$_{(p)}$ azimuthal direction. (c) RHEED intensity curves integrated within each rectangle window. (d)-(f) The results for the [100]$_{(p)}$ direction.
}
\label{fig2}
\end{center}
\end{figure}

Figure 2 shows typical RHEED images of the {\PSO} substrate and {\BIO} film along the [110]$_{(p)}$ and [100]$_{(p)}$ azimuth. {\BIO} films exhibit prominent Kikuchi lines, indicating high crystalline perfection, ensuring the quality of the photoemission spectra. The in-plane lattice mismatch to the {\PSO} substrate ($a_{p}=4.021$ {\AA}) induces only 0.2\% compressive strain and the RHEED streaks do not show a discernible shift resulting from any lattice relaxation (Figs. 2(c) and (f)).

\begin{figure}
\begin{center}
\includegraphics*[width=15cm]{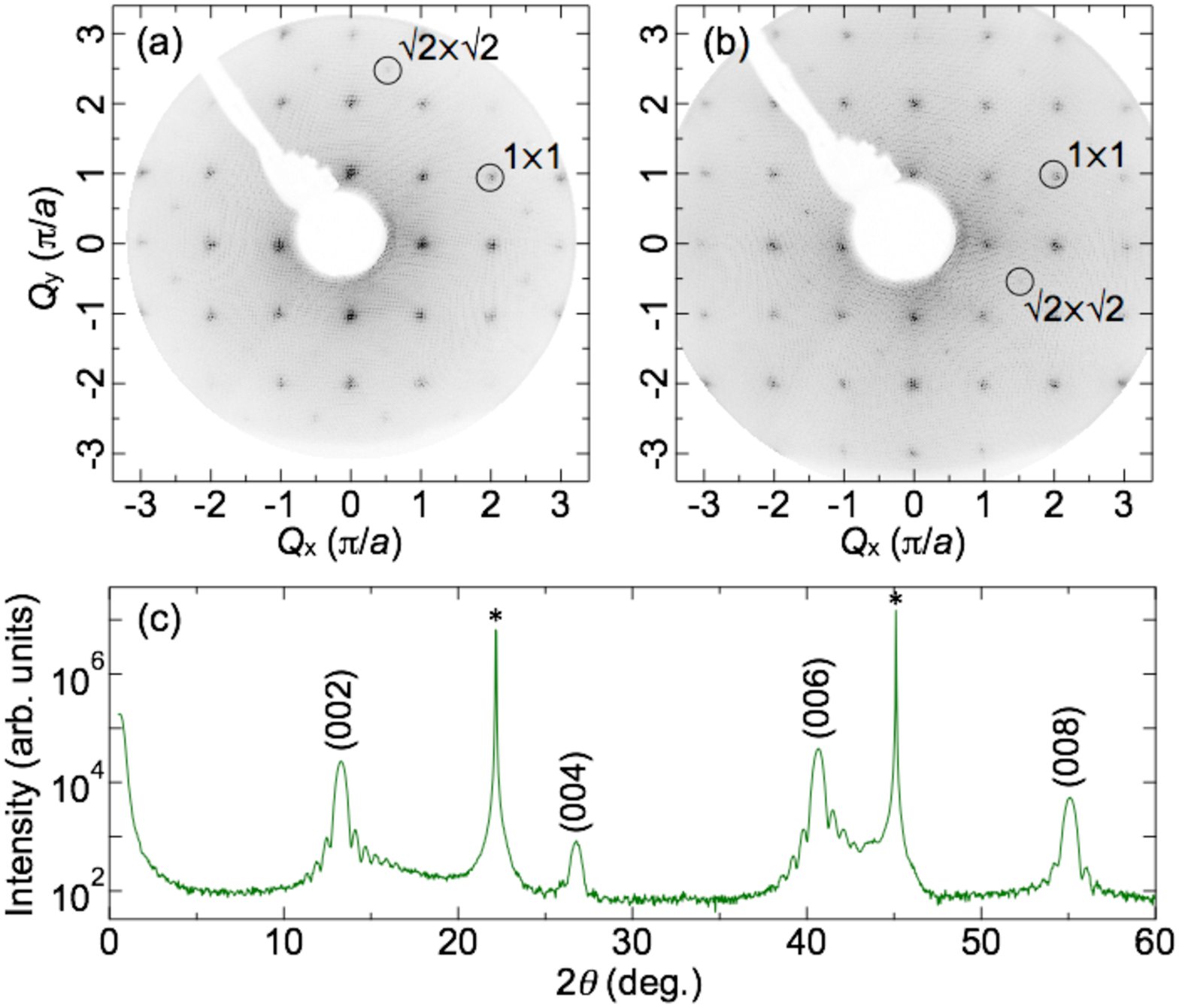}
\caption{(Color online) LEED images of a {\BIO} film, taken along the [001] direction with an electron energy of (a) 150 and (b) 200 eV. (c) XRD $\theta$--2$\theta$ scan of a 16 nm thick {\BIO} film grown on {\PSO} substrate. Substrate peaks are marked with an asterisk. 
}
\label{fig3}
\end{center}
\end{figure}

Exemplary LEED patterns are shown in Figs. 3(a) and (b), taken at normal incidence with beam energies of 150 and 200 eV. The sharp diffraction peaks indicate a well ordered surface crystal structure. The assumed tetragonal ($1\times 1$) diffraction peaks are indicated in the pattern, and weak Bragg peaks observed at $\sqrt{2}\times \sqrt{2}$ R45$^{\circ}$ relative to the $1\times 1$ peaks are likely a result of an in-plane surface reconstruction in {\BIO}, given the absence of such peaks in bulk x-ray measurements. The XRD $\theta$--2$\theta$ scan in Fig. 3(c) shows clear Kiessig fringes, indicating good surface smoothness, and a Nelson-Riley analysis of the peak positions gives an out-of-plane lattice constant of $c=13.34$ {\AA}, meaning films may be slightly elongated along the $c$-axis compared with bulk polycrystals ($a=4.030$ {\AA} and $c=13.333$ {\AA}) \cite{BIO1} due to the compressive in-plane strain.

\section{Results and discussion}

\begin{figure}
\begin{center}
\includegraphics*[width=15cm]{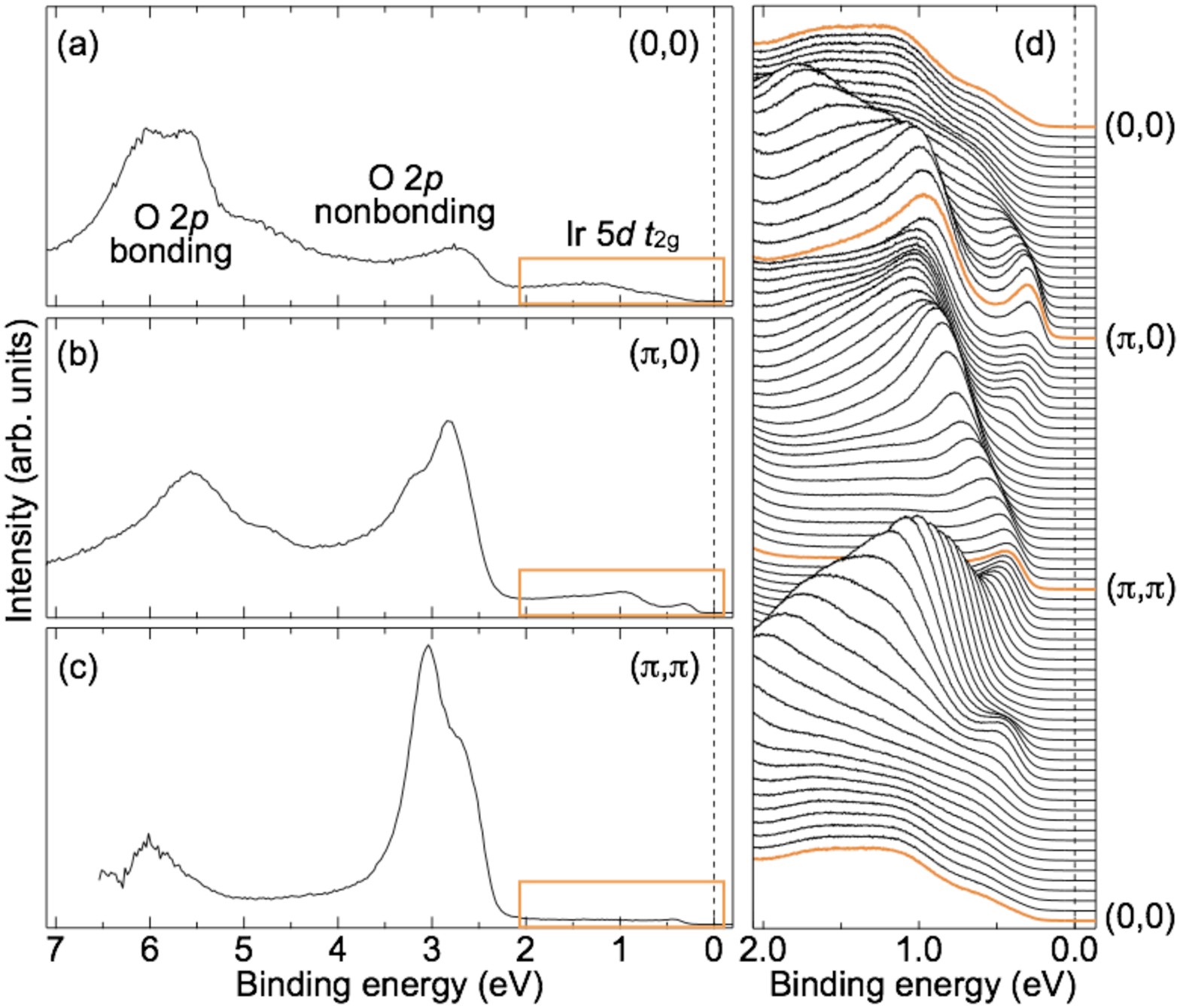}
\caption{(Color online) Valence band photoemission spectra at (a) $(0, 0)$, (b) $(\pi, 0)$, and (c) $(\pi, \pi)$ points for {\BIO}, respectively, taken at 100 K to prevent charging, showing Ir 5$d$ {\ttwog} bands and O 2$p$ bonding and nonbonding states at higher binding energies. (d) Magnified low-energy electronic structures along {\zerozero}--{\pizero}--{\pipi}--{\zerozero} high-symmetry lines of the tetragonal Brillouin zone.
}
\label{fig4}
\end{center}
\end{figure}

Figures 4(a)-(c) show valence band photoemission spectra at high-symmetry points for {\BIO} epitaxial films. By comparison to calculations,\cite{iridateSOI1, slater2, SIOcalc, iridateSC2TB1} the peaks near the Fermi level {\EF} can be assigned to the Ir 5$d$ {\ttwog} bands, while the spectral features between 2 and 7 eV can be ascribed to dominantly O 2$p$ states. As shown in the energy distribution curves (EDCs) in Fig. 4(d), low-energy dispersive features with clearly defined peaks are observed close to {\EF}, where the lowest-energy feature is located $\sim$0.3 eV below centered at the {\pizero} point, with another peak at the {\pipi} point only 0.4 eV below {\EF}. The full width at half maximum of the EDC peaks were typically $\sim$200 meV near the top of the valence band, with a lineshape well fit by a Gaussian. This indicates that the broad spectra are dominated by a manifold of states that involve multiple bosonic excitations, suggesting polaronic behavior, similar to that observed in the insulating parent cuprates,\cite{cuprateFC} as well as in the sister compound {\SIO}.\cite{SIOarpes3}

\begin{figure}
\begin{center}
\includegraphics*[width=14cm]{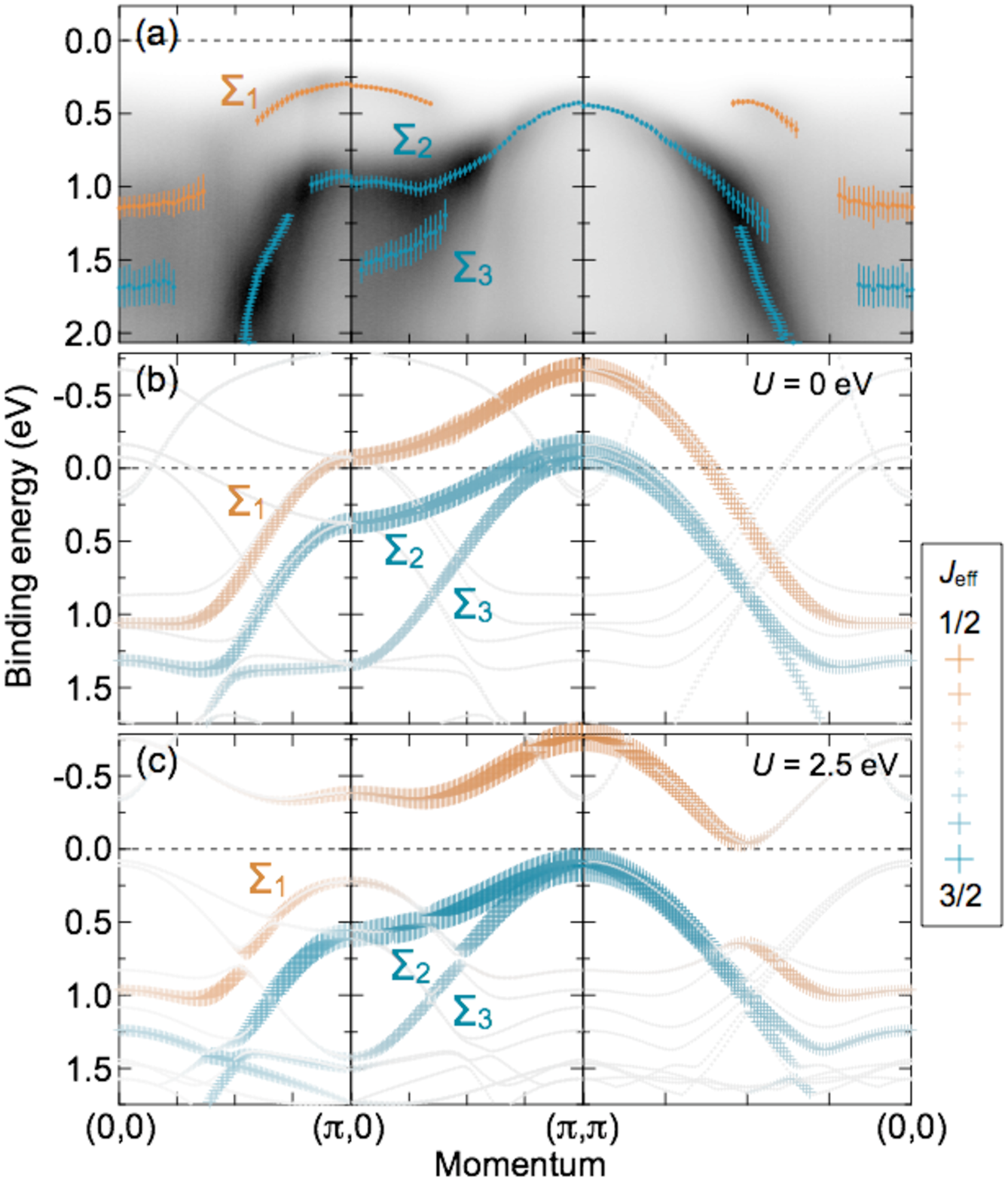}
\caption{(Color online) (a) Near-{\EF} ARPES spectral intensity plot along high-symmetry directions, taken at 100 K. Extracted EDC and MDC peak positions are overlaid with circles with error bars. DFT calculations including the spin-orbit coupling with $U$ of (b) 0 and (c) 2.5 eV. The $J_{\mathrm{eff}}$ character is calculated by projecting the eigenstates onto the {\Jhalf} and 3/2 bases in the atomic limit.
}
\label{fig5}
\end{center}
\end{figure}

In Figure 5, we compare the near-{\EF} ARPES spectra and the experimental dispersions ($\Sigma_{1}$--$\Sigma_{3}$) extracted from EDC and momentum distribution curve (MDC) fits, to our DFT calculations with spin-orbit coupling and an on-site Coulomb repulsion $U$. As shown in Fig. 5(a), along the {\pizero}--{\pipi} or {\halfhalf}--{\pipi} high symmetry directions, the half-filled $\Sigma_{1}$ band exhibits clear back-bending, indicating that a full gap opens along the {\pizero}--$(0, \pi)$ diagonal line, consistent with {\pipi} N{\' e}el antiferromagnetic order which is experimentally observed.\cite{iridateSOI2} The lowest energy states are at {\pizero}, where the peak position is at a binding energy of approximately 0.3 eV, consistent with earlier ARPES measurements on bulk {\SIO} and {\BIO} single crystals.\cite{iridateSOI1, BIObulk} Comparisons to our DFT calculations with $U$ of 2.5 eV (comparable to values reported in the literature for {\SIO} \cite{iridateSOI1, SIOcalc, SIOarpes2}) give qualitatively good agreement with the experimental dispersions, and indicate that the $\Sigma_{1}$ states which open a gap near {\pizero} are of predominately {\Jhalf} character. The $\Sigma_{2}$ and $\Sigma_{3}$ states near {\pipi} are observed to be very close in energy to the {\pizero} states, and in fact form the lowest lying states in the predicted valence band maximum at {\pipi}, at odds with experiment, but also observed in previous works on {\SIO}.\cite{iridateSOI1,SIOarpes2} Comparisons with the DFT calculations indicate that these states are of predominately {\Jthreehalf} character; the {\Jeff} character of these states do not appear to depend strongly on $U$ (Figs. 5(b) and (c)) or octahedral rotation angle, consistent with earlier work on {\SIO} by Martins {\emph{et al.}}.\cite{SIOcalc} This underscores the importance of low-lying {\Jthreehalf} states in the low energy physics of the layered iridates, and that the bandwidth of the {\Jhalf} and 3/2 states is substantially larger than their splittings. The faint dispersion especially around 0.5 eV centered at $(0, 0)$ is likely a folded feature of the corresponding $\Sigma_{2}$ band at {\pipi}, which is probably due to the $\sqrt{2} \times \sqrt{2}$ in-plane surface structural distortion observed in the LEED images (Figs. 3(a) and (b)). However, the effect of this reconstruction on the observed electronic structure appears much smaller than in {\SIO}, in that the intensity of the reflected shadow band around {\zerozero} is significantly weaker than at $(\pi,\pi)$, whereas in {\SIO} the intensity of these features are comparable. This may also imply that the octahedral rotation angles observed at the surface of the {\BIO} films are substantially smaller than those in {\SIO}.\cite{structmag1, structmag2}

Here it is meaningful to quantitatively compare the observed energy bands with the calculated ones. The experimentally extracted bandwidths (separation from band minimum to maximum) are $W_{\Sigma_{1}(0,0)-(\pi,0)}=0.9\pm 0.1$ eV, $W_{\Sigma_{2}(0,0)-(\pi,0)}=0.7\pm 0.1$ eV, $W_{\Sigma_{2}(\pi,0)-(\pi,\pi)}=0.5\pm 0.1$ eV, and $W_{\Sigma_{3}(\pi,0)-(\pi,\pi)}=1.1\pm 0.1$ eV. On the other hand, the corresponding values in the DFT calculations with $U$ of 2.5 eV are $0.74$ eV, $0.68$ eV, $0.48$ eV, and $1.16$ eV, respectively. Setting $U$ to zero only increases these bandwidths slightly to $1.13$ eV, $0.93$ eV, $0.54$ eV, and $1.48$ eV, but this weak renormalization reflects the relatively modest effective Coulomb interaction of the 5$d$ electrons, in contrast to the much larger bandwidth renormalization typically observed in the high-{\TC} cuprates. The other remarkable point is the energy difference between the {$\Sigma_{1}$} and {$\Sigma_{2}$} bands. Since this gap at the {$\Gamma$} point directly measures the spin-orbit coupling parameter as suggested by tight-binding models,\cite{iridateSC2TB1, Supplemental} the spin-orbit interaction in {\BIO} is experimentally determined to be $0.5\pm 0.1$ eV, which is also consistent to the reports on other iridate systems.\cite{iridateSOI1,Na2IrO3}

\begin{figure}
\begin{center}
\includegraphics*[width=10cm]{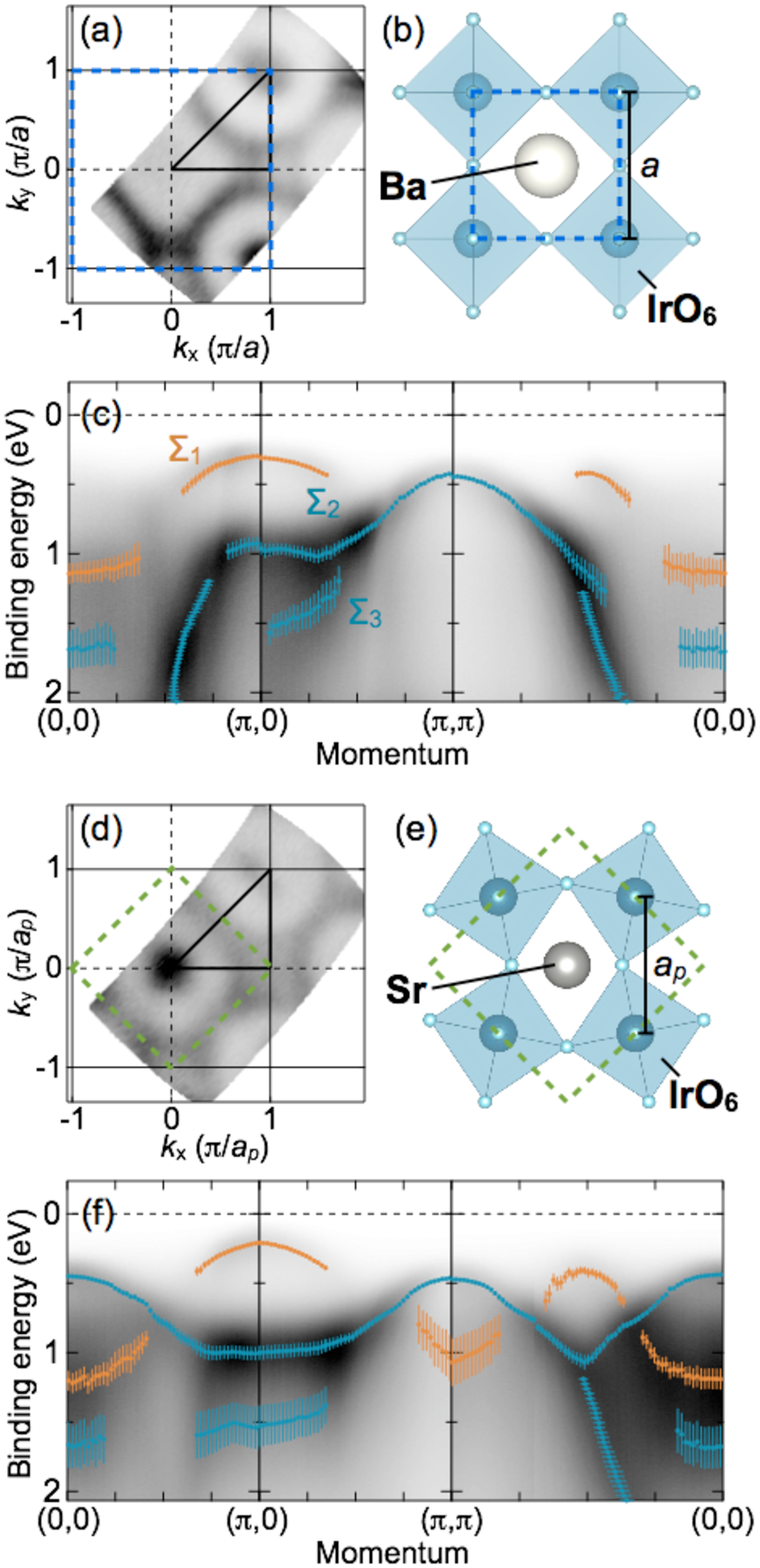}
\caption{(Color online) (a) Unsymmetrized isoenergy map at $0.4 \pm 0.005$ eV binding energy, (b) crystal structure projected on (001), and (c) $E$ versus $k$ spectra along high-symmetry lines for {\BIO}. Dashed squares in (a) and (b) indicate the in-plane Brillouin zone and unit cell, respectively. (d)-(f) The corresponding data from a {\SIO} thin film grown on LSAT is shown for comparison.
}
\label{fig6}
\end{center}
\end{figure}

Figure 6 shows a comparison of the measured low-energy electronic structures of {\BIO} and {\SIO}. (001) {\SIO} films were epitaxially grown on LSAT substrates by MBE, with the ARPES spectra consistent with previous measurements.\cite{iridateSOI1, SIOarpes2} As shown in Fig. 6(d), the Brillouin zone of {\SIO} is reduced by half in momentum space and the band structure is folded along the {\pizero}-$(0,\pi)$ and equivalent diagonal lines, reflecting the $\sqrt{2} \times \sqrt{2}$ in-plane distortion illustrated in Fig. 6(e). The folded bands are more clearly visible in the $E$ versus $k$ plots shown in Figs. 6(c) and (f). For example, the band dispersion seen centered at {\pipi} in {\BIO} can be clearly observed at the equivalent reconstructed position {\zerozero} in the case of {\SIO}. We interpret the weaker reconstruction observed in {\BIO} as an indication that the amount of octahedral rotation due to the surface reconstruction is significantly less than in {\SIO}. In addition, we find that the total bandwidth of the {\ttwog} bands are about 1.4 eV both in {\BIO} and {\SIO}. This suggests that the transfer integrals between the nearest-neighbor $d_{xy} $ ($d_{yz}$, $d_{zx}$) orbitals are roughly comparable.

\begin{figure}
\begin{center}
\includegraphics*[width=14.5cm]{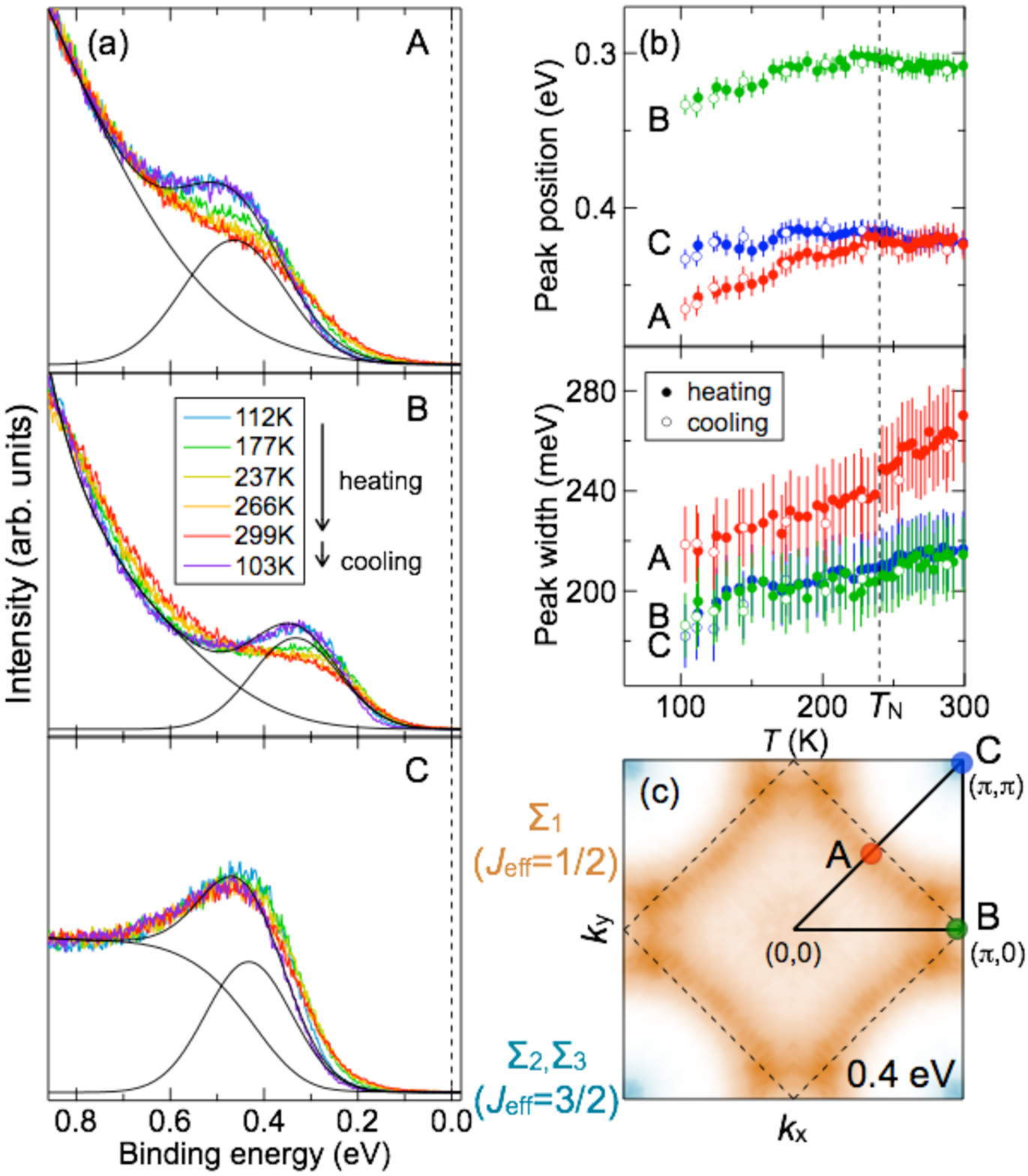}
\caption{(Color online) (a) Temperature-dependent EDCs at momentum positions A$\sim${\halfhalf}, B$\sim${\pizero}, and C$\sim${\pipi} as sketched in (c). The spectra were measured by cycling samples from 112 K to 299 K and then back to 103 K. Fitting results (one Gaussian curve and a background) for 103 K data are also represented by black curves. (b) Temperature evolution of the peak position (upper panel) and width (lower panel) for A--C. (c) Symmetrized isoenergy map at 0.4 eV in the full Brillouin zone, integrated within a $\pm 5$ meV window.
}
\label{fig7}
\end{center}
\end{figure}

Given that the {\Jhalf} and 3/2 states are nearly degenerate near {\EF}, we investigated their roles in the N{\' e}el transition by tracking their temperature dependence through {\TN}. Figure 7(a) shows temperature-dependent EDCs at the different momentum positions A--C sketched in Fig. 7(c). The spectra change reproducibly both in warming and cooling, ruling out the possibility of sample surface degradation. The effects of electrostatic charging and finite energy resolution were also experimentally precluded from affecting the data in any appreciable way. While peaks A (at {\halfhalf}) and B (at {\pizero}), primarily of {\Jhalf} character, clearly shift to higher binding energy with decreasing temperature below {\TN}, peak C (at {\pipi}), primarily of {\Jthreehalf} character, does not show any appreciable change in position. These observed spectral features can be suitably fit with a single Gaussian throughout the entire temperature range. In Fig. 7(b) we show the temperature dependence of the peak position and width quantitatively determined by our fits at the positions A--C. The peak positions at A and B on the {\Jhalf} band show an energy shift of about 40 meV in total below {\TN}, then abruptly stop shifting above {\TN}, while C on the {\Jthreehalf} shows negligible changes. Furthermore, the linewidth broadens continuously at a rate of $\sim0.2$ meV/K with increasing temperature, again expected from the thermal excitations of the bosonic dressing observed in polaronic systems such as {\SIO} and the undoped cuprates.\cite{SIOarpes3, cuprateFC} This unusual temperature dependence of the peak position suggests that only the {\Jhalf} band, which drives the insulator-metal transition, responds dramatically to the presence of long-range antiferromagnetism, while the {\Jthreehalf} states, which are nearby in energy, do not respond appreciably to the onset of the long-range antiferromagnetic order. The single particle excitation gap is directly enhanced by the presence of antiferromagnetic order, which would not be expected in simple Mott insulators, but suggestive of a Slater insulator, where the charge gap is directly tied to spin-density wave order. Structural origins of such a shift can be ruled out, as no sudden changes in the LEED patterns or in the crystal symmetry or lattice parameters have been observed either in {\BIO} or {\SIO} with temperature.\cite{BIO1, structmag1, structmag2}

The clear persistence of the gap and insulating behavior above {\TN} obviously indicates that the system does not behave as a simple Slater insulator. Furthermore, the nonmonotonic change of the observed {\Jhalf} states through {\TN} suggests the possibility that the effective Coulomb interactions may be enhanced below {\TN} accompanied by an ordering of the local moments. For example, in the antiferromagnetic Mott insulator {\LTO}, a similar nonmonotonic change of the Mott gap through {\TN} has been reported.\cite{LaTiO3opt}

\begin{figure}
\begin{center}
\includegraphics*[width=14.5cm]{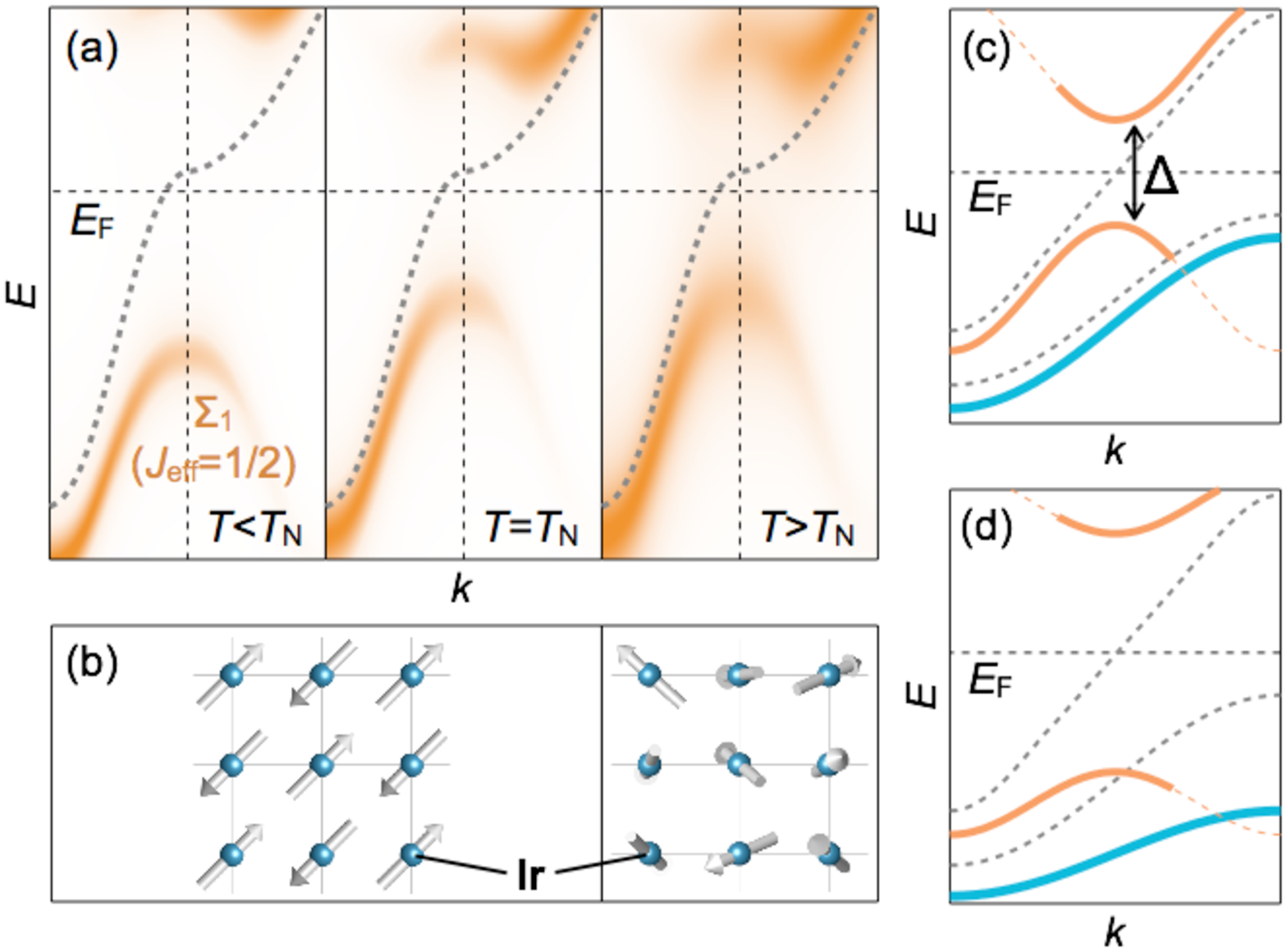}
\caption{(Color online) Schematic illustration of the temperature dependence of (a) the near-{\EF} band dispersion and (b) the in-plane spin state in {\BIO}. Dashed curves represent the original branch without including the electron correlation. Energy dispersions affected by (c) weak or (d) strong correlation are also shown.
}
\label{fig8}
\end{center}
\end{figure}

In Fig. 8(a) we illustrate schematically the temperature evolution of the gapped {\Jhalf} dispersion observed in {\BIO}. While the charge gap continues to soften when approaching {\TN}, it does not collapse above {\TN}, reflecting the robust correlated insulating state. On the other hand, the magnitude of the gap itself is relatively small and the bandwidth is found to be barely renormalized relative to the predictions from density functional theory, as schematically shown in Fig. 8(c). This is in stark contrast to conventional $3d$ transition metal oxide Mott insulators which feature a much larger charge gap and a more strongly renormalized bandwidth (Fig. 8(d)), a clear distinction between the parent insulating state of the layered iridates versus the cuprates. When the spin-orbit interaction is sufficiently strong, even a modest Coulomb repulsion is sufficient to push the nearly fully occupied {\Jthreehalf} band below {\EF} (blue), while splitting the half-filled {\Jhalf} band (orange).

\section{Conclusion}

In summary, we have investigated the electronic ground state and its temperature evolution in {\BIO} using a combination of reactive oxide molecular beam epitaxy and angle-resolved photoemission spectroscopy. A comparison between the experimental data and theoretical calculations have demonstrated the importance of the {\Jthreehalf} states for accurately describing the low-energy electronic structure. Our measurements have revealed that while the {\Jhalf} and 3/2 bands are barely renormalized relative to the band structure calculations, the two bands exhibit an unusual dichotomy in their behavior through the N{\' e}el transition. Only the split {\Jhalf} subband shows a substantial softening of the gap with increasing temperature approaching $T_{\mathrm{N}}$, suggesting an important role of the antiferromagnetic ordering in the formation of the insulating phase, while the gap itself remains robust well into the paramagnetic state. This surprising contrast in the behavior of the low-energy {\Jhalf} and 3/2 states should be crucial in developing accurate effective low-energy models to describe the parent insulating layered iridates and any emergent ground states, such as exotic superconductivity, that may be realized upon carrier doping. 


\begin{acknowledgments}
We gratefully acknowledge insightful discussions with J. W. Freeland, J. H. Lee, and T. Senthil. In particular, we would like to thank M. Grioni for helpful discussions regarding previously unpublished data in Ref. 24, which we became aware of during the preparation of this manuscript. This work was supported by the Air Force Office of Scientific Research (Grant No. FA9550-11-1-0033 and FA9550-12-1-0035), and the National Science Foundation through the MRSEC program (Cornell Center for Materials Research, DMR-1120296). M.U. acknowledges the support by JSPS Postdoctoral Fellowships for Research Abroad.
\end{acknowledgments}


\begin{thebibliography}{100}
\bibitem{SOIweyl1} X. Wan, A.M. Turner, A. Vishwanath, and S.Y. Savrasov, Phys. Rev. B \textbf{83,} 205101 (2011).
\bibitem{SOIweyl2} W. Witczak-Krempa and Y.B. Kim, Phys. Rev. B \textbf{85,} 045124 (2012).
\bibitem{SOItopo1} D. Pesin and L. Balents, Nat. Phys. \textbf{6,} 376 (2010).
\bibitem{SOItopo2} X. Zhang, H. Zhang, J. Wang, C. Felser, and S.-C. Zhang, Science \textbf{335,} 1464-1466 (2012).
\bibitem{iridateSC} F. Wang and T. Senthil, Phys. Rev. Lett. \textbf{106,} 136402 (2011).
\bibitem{iridateSC2TB2} H. Watanabe, T. Shirakawa, and S. Yunoki, Phys. Rev. Lett. \textbf{110,} 027002 (2013).
\bibitem{iridateSOI1} B. J. Kim, H. Jin, S. J. Moon, J. -Y. Kim, B. -G. Park, C. S. Leem, J. Yu, T. W. Noh, C. Kim, S. -J. Oh, J. -H. Park, V. Durairaj, G. Cao, and E. Rotenberg, Phys. Rev. Lett. \textbf{101,} 076402 (2008).
\bibitem{iridateSOI2} B. J. Kim, H. Ohsumi, T. Komesu, S. Sakai, T. Morita, H. Takagi, and T. Arima, Science \textbf{323,} 1329 (2009).
\bibitem{SIOopt1} S. J. Moon, H. Jin, W. S. Choi, J. S. Lee, S. S. A. Seo, J. Yu, G. Cao, T. W. Noh, and Y. S. Lee, Phys. Rev. B \textbf{80,} 195110 (2009).
\bibitem{slater1} D. Hsieh, F. Mahmood, D. H. Torchinsky, G. Cao, and N. Gedik, Phys. Rev. B \textbf{86,} 035128 (2012).
\bibitem{slater2} R. Arita, J. Kune{\v s}, A. V. Kozhevnikov, A. G. Eguiluz, and M. Imada, Phys. Rev. Lett. \textbf{108,} 086403 (2012).
\bibitem{slaterSTM} Q. Li, G. Cao, S. Okamoto, J. Yi, W. Lin, B. C. Sales, J. Yan, R. Arita, J. Kunes, A. V. Kozhevnikov, A. G. Eguiluz, M. Imada, Z. Gai, M. Pan, and D. G. Mandrus, Sci. Rep. \textbf{3,} 3073 (2013).
\bibitem{BIO1} H. Okabe, M. Isobe, E. Takayama-Muromachi, A. Koda, S. Takeshita, M. Hiraishi, M. Miyazaki, R. Kadono, Y. Miyake, and J. Akimitsu, Phys. Rev. B \textbf{83,} 155118 (2011).
\bibitem{BIO3RXS} S. Boseggia, R. Springell, H. C. Walker, H. M. R{\o}nnow, Ch. R{\" u}egg, H. Okabe, M. Isobe, R. S. Perry, S. P. Collins, and D. F. McMorrow, Phys. Rev. Lett. \textbf{110,} 117207 (2013).
\bibitem{BIOdoping} H. Okabe, M. Isobe, E. Takayama-Muromachi, N. Takeshita, and J. Akimitsu, Phys. Rev. B \textbf{88,} 075137 (2013).
\bibitem{BIO2} H. Okabe, N. Takeshita, M. Isobe, E. Takayama-Muromachi, T. Muranaka, and J. Akimitsu, Phys. Rev. B \textbf{84,} 115127 (2011).
\bibitem{IrOx} E. H. P. Cordfunke and G. Meyer, Recueil des Travaux Chimiques des Pays-Bas \textbf{81,} 495-504 (1962).
\bibitem{wien2k} P. Blaha, K. Schwarz, G. Madsen, D. Kvasnicka, and J. Luitz, Wien2k package, available at http:// www.wien2k.at.
\bibitem{PBE} J. P. Perdew, K. Burke, and M. Ernzerhof, Phys. Rev. Lett. \textbf{77,} 3865 (1996).
\bibitem{SIOcalc} C. Martins, M. Aichhorn, L. Vaugier, and S. Biermann, Phys. Rev. Lett. \textbf{107,} 266404 (2011).
\bibitem{iridateSC2TB1} H. Watanabe, T. Shirakawa, and S. Yunoki, Phys. Rev. Lett. \textbf{105,} 216410 (2010).
\bibitem{cuprateFC} K. M. Shen, F. Ronning, W. Meevasana, D. H. Lu, N. J. C. Ingle, F. Baumberger, W. S. Lee, L. L. Miller, Y. Kohsaka, M. Azuma, M. Takano, H. Takagi, and Z.-X. Shen, Phys. Rev. B \textbf{75}, 075115 (2007).
\bibitem{SIOarpes3} P. D. C. King, T. Takayama, A. Tamai, E. Rozbicki, S. M. Walker, M. Shi, L. Patthey, R. G. Moore, D. Lu, K. M. Shen, H. Takagi, and F. Baumberger, Phys. Rev. B \textbf{87,} 241106(R) (2013).
\bibitem{BIObulk} S. Moser, L. Moreschini, A. Ebrahimi, B. D. Piazza, M. Isobe, H. Okabe, J. Akimitsu, V. V. Mazurenko, K. S. Kim, A. Bostwick, E. Rotenberg, J. Chang, H. M. R{\o}nnow, and M. Grioni, New J. Phys. \textbf{16}, 013008 (2014).
\bibitem{structmag1} M. K. Crawford, M. A. Subramanian, R. L. Harlow, J. A. Fernandez-Baca, Z. R. Wang, and D. C. Johnston Phys. Rev. B \textbf{49,} 9198 (1994).
\bibitem{structmag2} Q. Huang, J. L. Soubeyroux, O. Chmaissem, I. Natali Sora, A. Santoro, R. J. Cava, J. J. Krajewski, and W. F. Peck Jr., J. Solid State Chem. \textbf{112,} 355 (1994).
\bibitem{Supplemental} See Supplemental Material at [URL will be inserted by publisher] for more information regarding tight binding calculation.
\bibitem{Na2IrO3} R. Comin, G. Levy, B. Ludbrook, Z.-H. Zhu, C. N. Veenstra, J. A. Rosen, Y. Singh, P. Gegenwart, D. Stricker, J. N. Hancock, D. van der Marel, I. S. Elfimov, and A. Damascelli, Phys. Rev. Lett. \textbf{109,} 266406 (2012).
\bibitem{SIOarpes2} Q. Wang, Y. Cao, J. A. Waugh, S. R. Park, T. F. Qi, O. B. Korneta, G. Cao, and D. S. Dessau, Phys. Rev. B \textbf{87,} 245109 (2013).
\bibitem{LaTiO3opt} P. Lunkenheimer, T. Rudolf, J. Hemberger, A. Pimenov, S. Tachos, F. Lichtenberg, and A. Loidl, Phys. Rev. B \textbf{68,} 245108 (2003).

\end{thebibliography}
\end{document}